\documentclass[%
 aip,
 amsmath,amssymb,
preprint,%
]{revtex4-2}

\usepackage{graphicx}
\usepackage{dcolumn}
\usepackage{bm}

\usepackage[utf8]{inputenc}
\usepackage[T1]{fontenc}
\usepackage{mathptmx}
\usepackage{etoolbox}
\usepackage[capitalize]{cleveref}
\newcommand{\figscale}{0.9}

\makeatletter
\def\@email#1#2{%
 \endgroup
 \patchcmd{\titleblock@produce}
  {\frontmatter@RRAPformat}
  {\frontmatter@RRAPformat{\produce@RRAP{*#1\href{mailto:#2}{#2}}}\frontmatter@RRAPformat}
  {}{}
}%
\makeatother
\begin{document}


	\title{Photonic next-generation reservoir computer based on distributed feedback in optical fiber}

\author{Nicholas Cox}
\email{nicholas.a.cox34.ctr@us.navy.mil} 
\author{Joseph Murray}
\author{Joseph Hart}
\author{Brandon Redding}

\affiliation{U.S. Naval Research Laboratory, 4555 Overlook Ave, SW, Washington, DC 20375, USA}

\date{\today}

\begin{abstract}
		Reservoir computing (RC) is a machine learning paradigm that excels at dynamical systems analysis. Photonic RCs, which perform implicit computation through optical interactions, have attracted increasing attention due to their potential for low latency predictions. However, most existing photonic RCs rely on a nonlinear physical cavity to implement system memory, limiting control over the memory structure and requiring long warm-up times to eliminate transients. In this work, we resolve these issues by demonstrating a photonic next-generation reservoir computer (NG-RC) using a fiber optic platform. Our photonic NG-RC eliminates the need for a cavity by generating feature vectors directly from nonlinear combinations of the input data with varying delays. Our approach uses Rayleigh backscattering to produce output feature vectors by an unconventional nonlinearity resulting from coherent, interferometric mixing followed by a quadratic readout. Performing linear optimization on these feature vectors, our photonic NG-RC demonstrates state-of-the-art performance for the observer (cross-prediction) task applied to the R\"ossler, Lorenz, and Kuramoto-Sivashinsky systems. In contrast to digital NG-RC implementations, this scheme is easily scalable to high-dimensional systems while maintaining low latency and low power consumption.
\end{abstract}

\maketitle
\begin{quotation}
	Reservoir computing (RC) is a machine learning technique that has shown great success in chaotic systems analysis. For RCs implemented in digital electronics, the power consumption and computational latency are prohibitive for time and energy-sensitive applications. These limitations can be overcome by performing computations in the analog domain using a physical system. In particular, photonic RCs are especially promising as they exploit the high data encoding rates and parallelism of optical signals. This work demonstrates a new photonic reservoir computing architecture that uses Rayleigh backscattering in optical fiber as the main computational engine. In contrast to previous photonic RCs, which form a reservoir through a complicated feedback system, we achieve comparable computational accuracy using a simple feedforward approach that implements a virtual reservoir by applying a random mixing transformation to a fixed number of consecutive time samples at a given time step. We employ this architecture to create a time series observer which cross-predicts unknown state variables of a multi-dimensional chaotic system from a subset of measurable states. We optimize our experimental system using the additional degree of memory control provided by the virtual reservoir, achieving state-of-the-art performance on a series of benchmark observer tasks.
\end{quotation}

\section{Introduction}

Recurrent neural networks (RNN) are a class of artificial neural networks especially suited for dynamical systems analysis. RNNs derive their computational power from  the memory generated by an optimized feedback loop between neuron layers. While this architecture has shown considerable success, the learning process often suffers from vanishing or exploding gradients \cite{rumelhartLearningRepresentationsBackpropagating1986}. Reservoir computing (RC) is an RNN subclass developed around a recurrent feedback network connected with fixed random weights. RCs avoid the training pitfalls of standard RNNs because they require optimization of only a single output layer that maps the reservoir state to a desired output. The idea of employing a physical system as a reservoir dates back to the conception of RC in the form of echo state networks \cite{jaegerHarnessingNonlinearityPredicting2004} and liquid state machines \cite{maassRealTimeComputingStable2002}. Shortly after, researchers realized RCs using water buckets \cite{fernandoPatternRecognitionBucket2003}, VLSI chips \cite{schurmannEdgeChaosComputation2004}, and electronic feedback loops \cite{appeltantInformationProcessingUsing2011}. 

Optics serves as a particularly promising platform for physical reservoirs because the high modulation bandwidth and multiplexing capability of light enable high-speed low-power computational inference. Paquot et al. \cite{paquotOptoelectronicReservoirComputing2012} and Larger et al. \cite{largerPhotonicInformationProcessing2012} developed the first photonic RCs using a delay-line architecture originally proposed for electronic RCs \cite{appeltantInformationProcessingUsing2011}. This design, which remains the most commonly investigated approach \cite{chemboMachineLearningBased2020,martinenghiPhotonicNonlinearTransient2012,brunnerTutorialPhotonicNeural2018,hartDelayedDynamicalSystems2019,duportVirtualizationPhotonicReservoir2016,duportFullyAnaloguePhotonic2016,dejonckheereAllopticalReservoirComputer2014,carrollTimeShiftsReduce2022,vatinExperimentalReservoirComputing2019,argyrisPhotonicMachineLearning2018,nguimdoPredictionPerformanceReservoir2017,sorianoOptoelectronicReservoirComputing2013,ortinUnifiedFrameworkReservoir2015,antonikOnlineTrainingOptoElectronic2017,harkhoeDemonstratingDelaybasedReservoir2020,sozosReservoirComputingBased2021,danilenkoImpactFilteringPhotonic2023,donatiMicroringResonatorsExternal2022,hulserRoleDelaytimesDelaybased2022,abdallaMinimumComplexityIntegrated2023}, uses delayed feedback following a single nonlinear transformation to define time-multiplexed virtual nodes. Delay-line RCs have consistently demonstrated state-of-the-art performance for tasks such as speech classification \cite{largerHighSpeedPhotonicReservoir2017},  time series prediction \cite{liangRealtimeRespiratoryMotion2023}, and distortion compensation for optical communications  \cite{argyrisPhotonicMachineLearning2018}, but the virtual node structure proves challenging to scale for the processing of high-dimensional data. Multi-channel methods have been recently explored to improve scalability \cite{ takanoCompactReservoirComputing2018,guoFourchannelsReservoirComputing2019,ortinReservoirComputingEnsemble2017,houPredictionClassificationPerformance2019,keuninckxRealtimeAudioProcessing2017,yueReservoirComputingBased2021,yangOpticalNeuromorphicComputing2022,bogrisFabryPerotLasersEnablers2021}, but they come at the cost of increased complexity. An alternative structure involving the coherent mixing of spatial electromagnetic field profiles in free-space \cite{paudelClassificationTimedomainWaveforms2020,ashnerPhotonicReservoirComputer2021,antonikLargeScaleSpatiotemporalPhotonic2020,rafayelyanLargeScaleOpticalReservoir2020,buenoReinforcementLearningLargescale2018,sunadaUsingMultidimensionalSpeckle2020}  aims to overcome this limit on the maximum number of nodes, but the computation speed is severely limited by the update rate of spatial light modulators or digital micromirror devices.

A common feature of both the standard delay-line and free-space designs is the existence of a nonlinear feedback loop that dictates the evolution of an internal reservoir state. Performing a read-out operation on this internal state yields an output feature vector. This feature vector is finally mapped to the desired output by linear optimization using a training data set. As an alternative to nonlinear feedback, Refs. \cite{vandoorneExperimentalDemonstrationReservoir2014, vinckierHighperformancePhotonicReservoir2015, nakajimaCoherentlyDrivenUltrafast2018, nakajimaScalableReservoirComputing2021,butschekPhotonicReservoirComputer2022} demonstrate the feasibility of high-performance RC using a fully linear optical reservoir followed by a nonlinear readout (e.g. from the photodetector response). Recently, Gauthier et al. \cite{gauthierNextGenerationReservoir2021} introduced next generation reservoir computers (NG-RC) as a further simplification of the linear reservoir concept. An NG-RC removes the reservoir entirely by explicitly constructing output feature vectors from various nonlinear combinations of data from current and previous time steps.  Although NG-RCs have shown impressive performance on a variety of tasks with reduced training and warm-up time \cite{barbosaLearningSpatiotemporalChaos2022, gauthierLearningUnseenCoexisting2022, kentControllingChaoticMaps2024}, the generation of nonlinear feature vectors is computationally intensive for high-dimensional systems or systems with high-order nonlinearities. While photonics might be used to accelerate this process, creating feature vectors of the form proposed in Ref. \cite{gauthierNextGenerationReservoir2021} on a photonic platform would require a complex optical routing scheme and the ability to implement a variety of nonlinear transforms. 

In this work, we show that state-of-the-art performance on benchmark tasks is achievable using a different form of feature vector created by applying random nonlinear projections to the input data. As a proof-of-concept, we present a photonic NG-RC using a simple fiber optic platform that can efficiently perform these projections on high-dimensional data with low latency and low power consumption \cite{reddingFiberOpticComputing2024}. The photonic NG-RC operates by encoding an input feature vector containing current and past data points onto the phase of an optical pulse, injecting the pulse into standard single-mode fiber, and recording the Rayleigh backscattering pattern measured by a photodetector. The net nonlinear transformation, provided by coherent complex-domain mixing followed by a quadratic readout, is a temporal analog to the one performed in the spatial domain in Refs. \cite{paudelClassificationTimedomainWaveforms2020, ashnerPhotonicReservoirComputer2021, antonikLargeScaleSpatiotemporalPhotonic2020, rafayelyanLargeScaleOpticalReservoir2020}. Comparing our fiber-optic design to this spatial approach, we find  advantages in that (1) our system is not bottle-necked by the speed limitations of a spatial light modulator and (2) the NG-RC framework removes the need for nonlinear optical-electronic-optical feedback. Additionally, when compared to delay line RCs, we find better scalability to high-dimensional data analysis while maintaining the benefits of digital NG-RCs: short warm up time, reduced training duration, and straightforward optimization and control of hyperparameters such as memory design. Exploiting this control over memory structure, we introduce a novel mask procedure that transforms elements of the input feature vectors at different time delays. We find that tailoring this mask can improve performance of the photonic NG-RC, especially for the analysis of high-dimensional data. 

We illustrate the versatility of the photonic NG-RC by applying it to a range  of time series observer tasks \cite{luReservoirObserversModelfree2017}. The role of an observer is to examine a limited subset of state variables in a multi-dimensional dynamical system and cross-predict the remaining states \cite{zimmermannObservingSpatiotemporalDynamics2018, cunilleraCrosspredictingDynamicsOptically2019, neofotistosMachineLearningObservers2019, barmparisRobustPredictionComplex2020, doanLearningHiddenStates2020, pammiExtremeEventsPrediction2023}. This cross-prediction task is important for both real-world systems (where some quantities are unmeasurable but crucial for control or decision-making purposes) and as a benchmark RC task (since successful cross-predictions for chaotic systems relies on a delay-coordinate embedding \cite{packardGeometryTimeSeries1980} of the attractor that is highly sensitive to the memory properties). In this work, we test the photonic NG-RC observer performance on the chaotic R\"ossler \cite{rosslerEquationContinuousChaos1976}, Lorenz \cite{lorenzDeterministicNonperiodicFlow1963}, and Kuramoto-Sivashinsky (KS) \cite{kuramotoDiffusionInducedChaosReaction1978,sivashinskyNonlinearAnalysisHydrodynamic1977} systems. For each system, we optimize the memory by varying the number of delay samples present in the input feature vector as well as the mask function applied to them. We use the R\"ossler equations, characterized by the simplest chaotic attractor, to present initial results and demonstrate the ability to achieve high performance with relatively few training points. For the Lorenz system, we study the effect of varying the system sampling rate after completing the initial memory optimization. Finally, we use the spatiotemporal KS system to highlight the ability of the fiber NG-RC to analyze high-dimensional data. Finding that the photonic NG-RC achieves state-of-the-art performance for each task, we then examine the potential of our design to perform these predictions with low latency and low power consuption. We compare the power consumption to that of digital electronic NG-RCs to identify tasks for which the photonic system is best suited.

\section{Experimental design and operation} 

\begin{figure}[!h]
	\centering
	\includegraphics[scale=\figscale]{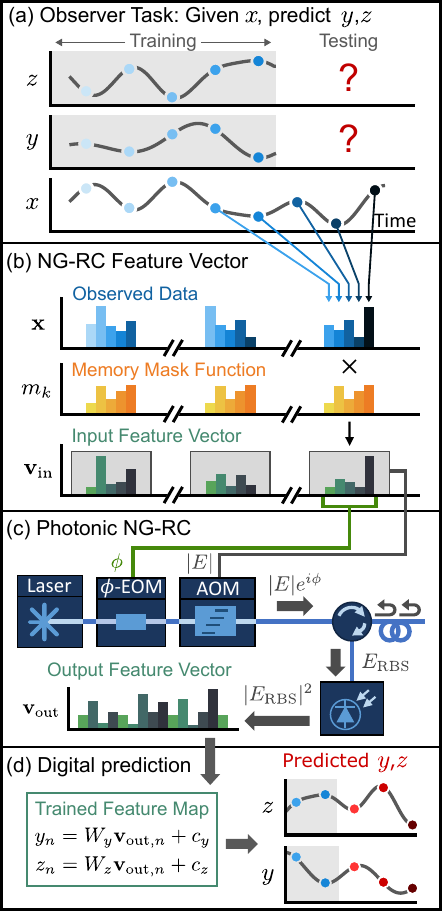}
	\caption{An overview of the photonic NG-RC system, detailing (a) the problem definition, (b) the construction of input feature vectors, (c) the experimental fiber-optic setup, and (d) the final prediction step in digital electronics.}
	\label{fig:RC_design}
\end{figure}

The photonic NG-RC takes an input feature vector comprising a fixed number of consecutive time-delay samples and performs a nonlinear random projection to generate an output feature vector in higher-dimensional space. This process is depicted in \cref{fig:RC_design} for an observer task in which the $x$ component of a chaotic time series is known and the objective is to cross-predict $y$ and $z$. An input feature vector is first constructed in software by concatenating the current value of $x$ with $K-1$ previous time samples. Unlike a traditional RC, for which the cavity lifetime dictates the system memory, the NG-RC memory is easily controlled by adjusting the number of time steps included in the input feature vector. In this work, we introduce a memory mask function, $m_k$, to scale the “Observed Data” (\cref{fig:RC_design}(b)) in the input feature vector. The mask provides an additional degree of control in the photonic NG-RC, which we later find to be essential for optimizing the system performance. The input feature vector, $\mathbf{v}_{\text{in},n}$, at the $n^{\text{th}}$ time step is defined to be
\begin{equation}
	\mathbf{v}_{\text{in},n} =\mathbf{x}_n \oplus m_1  \mathbf{x}_{n-1} \oplus \cdots \oplus m_{K-1}  x_{n-(K-1)}.
\end{equation}
The symbol $\oplus$ is the direct sum (concatenation) operator, $\mathbf{x}_n$ is a vector composed of each observed state variable at $n^\text{th}$ time step (in the case shown in \cref{fig:RC_design} with a single observed variable, $\mathbf{x}_n= x_n$), and $m_k$ is the multiplicative mask applied to the NG-RC memory. Here, we considered three masks: (1) the unit mask $m_k=1$, (2) the decaying mask $m_k = \beta^k$ (with $\beta = 0.8$), and (3) the random mask comprising pseudorandom numbers between 0 and 1. In each case, the mask function remains fixed for each sample step $n$.

As shown in \cref{fig:RC_design}(c), the photonic NG-RC encodes input vectors into the phase of an optical carrier using an electro-optic phase modulator ($\phi$-EOM). We used phase modulation because it allows for a more natural encoding of positive and negative values compared to amplitude modulation which requires a voltage offset to accommodate negative values. An amplitude modulator (an acousto-optic modulator, AOM, was used in this work) then carves a pulse surrounding the phase modulated region. Note that the pulse duration was chosen to be slightly longer than the phase-modulated region to provide a fixed phase reference in each pulse. This encoded pulse was injected into a stretch of standard single  mode fiber and the Rayleigh backscattered (RBS) light was measured by a photodetector. The resulting voltage, proportional to the RBS pattern irradiance, was finally digitized using an analog-to-digital converter (ADC) to yield the output feature vector. The length of this output feature vector is proportional to the scattering fiber length, providing our design with a simple method of tailoring the output dimensionality to a specific problem.

We can model the above procedure to understand the nonlinear random projection performed by the photonic NG-RC. To begin, input feature vectors are encoded into the phase modulator driving voltage
\begin{equation}\label{eq:Vn}
	V_{\text{in},n}(t) \approx \sum_{\mu} v_{\text{in},n}^\mu \text{rect} \left(\frac{t-\mu\tau}{\tau}\right),
\end{equation}
where $v^\mu_{\text{in},n}$ is the $\mu$ component of $\mathbf{v}_{\text{in},n}$ and $\tau$ is the duration of a single rectangular pulse encoding an input element. After the AOM defines the envelope $|E(t)|$, the modulated optical signal passes into the scattering fiber. Measuring the subsequent RBS irradiance using a photodector yields the voltage
\begin{equation}\label{eq:trans}
	V_{\text{out},n}(t) \approx \left| |E(t)|e^{i (\omega t + \pi V_{\text{in},n}(t)/V_{\pi})}\ast r_{\text{RBS}}(t)\right|^2,
\end{equation}
where $\omega$ is the optical angular frequency, $V_\pi$ is the modulator voltage to impart a phase shift of $\pi$, $(*)$ is the convolution operator, and $r_{\text{RBS}}(t)$ is the Rayleigh backscattering response of the optical fiber (See SI for a full mathematical description). The output vector is finally formed by sampling \cref{eq:trans} at a fixed rate $t_s$:
\begin{equation}
	v^\mu_{\text{out},n} \approx V_{\text{out},n}(\mu t_s).
\end{equation}
We finally see that the overall transformation $\mathbf{v}_{\text{in},n} \rightarrow \mathbf{v}_{\text{out},n}$ is nonlinear due to the voltage relationship in \cref{eq:trans} including a convolution followed by quadratic readout.

As in a traditional reservoir observer, the NG-RC operation assumes there is an initial training period during which all state variables of a dynamical system are available. This training period is used to optimize a output weight matrix ($W$) containing one row for each predicted variable. After the training period, the output weights are held fixed and used to predict the unknown state variables ($y$,$z$ in \cref{fig:RC_design}) during the testing phase (e.g. $y_n=W_y \mathbf{v}_{\text{out},n}+c_y$, where $c_y$ is a DC offset). In this work, the prediction stage was performed via digital electronics after digitizing the output feature vector. In the future, the weight matrix could be applied in the analog photonic domain using the technique developed in Refs. \cite{duportFullyAnaloguePhotonic2016,smerieriAnalogReadoutOptical2012} to minimize the system latency. 

Experimentally, we constructed the photonic NG-RC using the basic architecture shown in \cref{fig:RC_design}(c). A narrow-linewidth ($\sim$kHz) seed laser operating at a wavelength of $\sim$1550 nm is directed to an electro-optic phase modulator to encode the input feature vector. Using an arbitrary waveform generator (AWG), each element of the input feature vector $\text{v}_{\text{in},n}$ is encoded into the peak voltage of a $\tau=10$ ns rectangular pulse. These $\phi$-EOM input voltages are scaled so that the maximum and minimum values over all input vectors are $\pm V_\pi$. The length of the phase encoded input vector is $KQ \tau$, where $K$ is the memory length (i.e. the number of time steps included in the input feature vector) and $Q$ is the number of observed input dimensions ($Q=1$  in the example shown in \cref{fig:RC_design}). The phase-encoded light is then directed to an AOM which forms an amplitude envelope of width $(KQ+6)\tau$ centered on the phase-encoded signal. The buffer time of 6$\tau$ allows for a portion of the unmodulated source to act as a zero phase-shift reference (experimentally, we found that the NG-RC performance is relatively insensitive to the length of this buffer within the range of 2$\tau$ to 20$\tau$). The pulse was then amplified using an erbium doped fiber amplifier (EDFA) and coupled through a circulator to a section of single mode fiber of length $L=500$ m. The RBS light was directed by the circulator to a second EDFA before reaching a 120 MHz bandwidth photodetector. The detected signal was digitized at 1 GS/s.  In this proof-of-concept demonstration, the first EDFA was used to compensate for insertion loss through the EOM and AOM, while the second EDFA was used to mitigate the effect of detector and digitizer noise. In the future, using a higher power seed laser, enhanced Rayleigh backscattering fiber, lower insertion loss modulators, and a lower noise detector  could obviate the need for the EDFAs. 

The total duration of the RBS pattern, and hence the output feature vector, was $\tau_{\text{RBS}} =2L/(c/n_g)+KQ\tau \approx 5$ $\mu$s, where $c$ is the vacuum speed of light and $n_g$ is the group index. This entire process was repeated for each input time step in a data set at a repetition period of $T=6$ $\mu$s (set to be slightly longer than $\tau_\text{RBS}$). In each experiment presented below, we sent a total of $N$ input feature vectors including information from $N$ consecutive samples of a dynamical system. After retrieving all output feature vectors, we partitioned into a set of $N_{\text{train}}$ training points followed by $N_{\text{test}}$ testing points. The output weight matrix ($W$) was then computed offline by ridge regression and used to cross-predict output variables in the test phase. 

To mitigate experimental noise, we repeated the measurements 25 times (R\"ossler, Lorenz) or 5 times (KS) and averaged the resulting RBS patterns to obtain a lower noise output feature vector (see SI for details on the effect of averaging on the prediction accuracy and measurement SNR). Although the RBS pattern was sampled at 1 GHz , the temporal correlation width of the RBS light is set by the $\tau=10$ ns encoding bandwidth \cite{mermelsteinRayleighScatteringOptical2001}. In this work, we performed a 5-point moving mean filter and down-sampled the measured pattern at half the correlation width, or a rate of 200 MS/s, retaining some partially correlated information. This resulted in an output feature vector of length $D \approx \tau_{\text{RBS}}/(\tau/2)=1000$. 

\section{R\"ossler System}

\begin{figure}[h!]
	\centering
	\includegraphics[scale=\figscale]{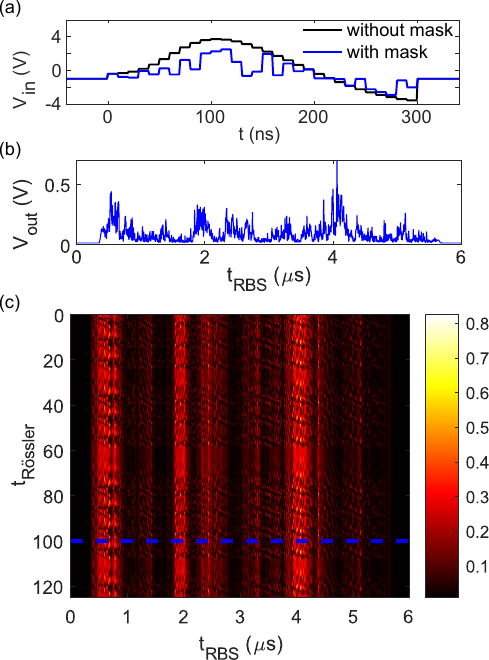}
	\caption{Experimental data for the R\"ossler observer. a) the phase modulator input voltage that encodes the input vector $\mathbf{v}_{\text{in},n}$  at $n=800$ ($t_{\text{R\"ossler}} = 100$). b) the voltage output proportional to the RBS irradiance that is downsampled to form the output vector $\mathbf{v}_{\text{out},n}$ at $n=800$. c) a plot of the first 1000 retrieved speckle patterns, with the dotted blue line marking the location of the pattern shown in part b.}
	\label{fig:rossler_fig1}
\end{figure}

We first applied the photonic NG-RC to analyze the R\"ossler system, which is a classic chaotic system often used to benchmark RC performance. The R\"ossler system is governed by the coupled differential equations \cite{rosslerEquationContinuousChaos1976} 
\begin{align}\label{eq:rossler}
	x_t &= -y - z \nonumber \\
	y_t &= x + ay \nonumber \\
	z_t &= b + z(x-c),
\end{align}
where the subscript $t$ denotes a derivative with respect to time. We use the standard parameters $a=0.5$, $b=2.0$ and $c=4.0$, which are known to exhibit chaos \cite{luReservoirObserversModelfree2017}. In this section, we train the fiber NG-RC to use $x$ as an input to cross-predict the variables $y$ and $z$. Ground truth data was generated by randomly selecting a set of initial conditions  and integrating \cref{eq:rossler} with a step size $\Delta t_{\text{R\"ossler}} = 0.125$ using the fourth order Runge-Kutta method. A typical input feature vector with memory length $K=30$ and a random input mask is shown in \cref{fig:rossler_fig1}a, and the resulting output feature vector produced by the fiber optic platform appears in \cref{fig:rossler_fig1}b. This process was repeated for $N=3000$ time steps, of which the first 1000 are shown in \cref{fig:rossler_fig1}c.

\begin{figure*}[!htb]
		\includegraphics[scale=\figscale]{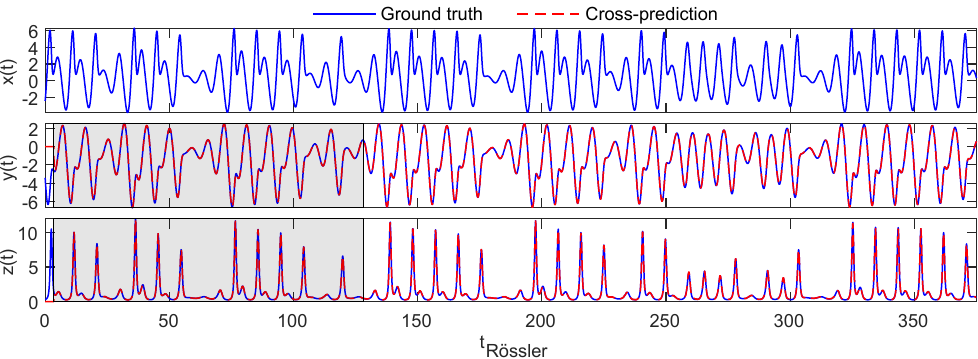}
	\caption{Output of the Photonic NG-RC observer (red dashed) with $K=30$ and a random mask compared to ground truth (blue solid) for $N=3000$ samples of the R\"ossler system. The grey shaded region marks the training set of $N_{\text{train}}=1000$ points following a $K-1 = 29$-step warm-up time. The errors for the data shown are $\text{NRMSE}_y=1.23 \times 10^{-2}$, $\text{NRMSE}_z=1.89 \times 10^{-2}$.}
	\label{fig:rossler_single}
\end{figure*}

We allocated the first $N_{\text{train}}=1000$ points (corresponding to $t_{\text{R\"ossler}} =N_{\text{train}}\Delta t_{\text{R\"ossler}} =125$) for training and used ridge regression to calculate the weight matrix that minimized errors in the cross-predicted values of $y$ and $z$. We then tested the system performance on the remaining $N_{\text{test}}=2000$ data points. As shown in \cref{fig:rossler_single}, the NG-RC accurately predicted the $y$ and $z$ coordinates of the R\"ossler system after the training period ($t_{\text{R\"ossler}} > 125$) using an input feature vector  with memory of length $K=30$ and a random memory mask. 
To quantitatively evaluate the photonic NG-RC performance, we computed the normalized root mean square error (NRMSE) for the variable $y$ (and similarly for $z$) during the testing phase ($n>N_{\text{train}}$), as
\begin{equation}
	\text{NRMSE}_y = \frac{1}{\sigma}\sqrt{\sum_n \left(y_n - \hat{y}_n\right)^2},
\end{equation}
where $\hat{y}_n$ is the RC prediction and $y_n$ is the ground-truth value at time step $n$. The error is normalized by the standard deviation, $\sigma$, of the ground truth curve. For ease of comparison, we averaged the NRMSE for outputs y and z, defining $\text{NRMSE}=(\text{NRMSE}_y+ NRMSE_z)/2$. For $K=30$ case shown in \cref{fig:rossler_single}, we found a NRMSE of $1.56\times 10^{-2}$ ($\text{NRMSE}_y=1.23 \times 10^{-2}$, $\text{NRMSE}_z=1.89 \times 10^{-2}$). 

\begin{figure}[!htb]
	\centering
	\includegraphics[scale=\figscale]{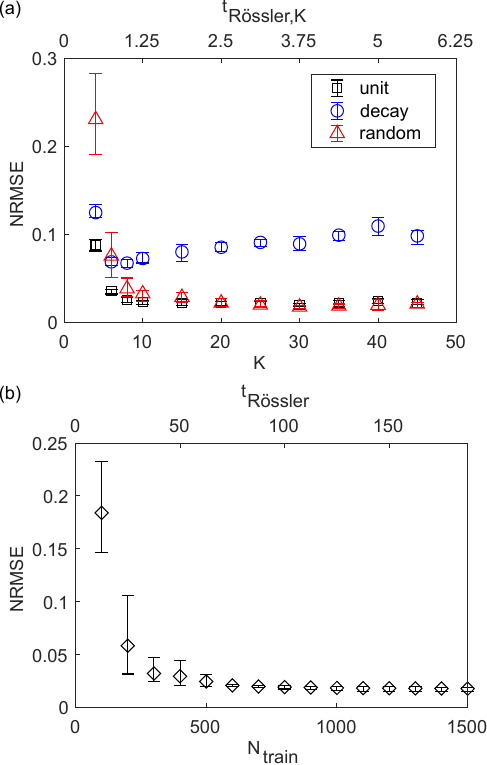}
	\caption{a) NRMSE versus memory length $K$/duration $t_{\text{R\"ossler},K}$ and mask type for the Rössler observer. b) NRMSE for the Rössler system versus number of training points ($N_\text{train}$) using a memory with $K=30$ and a random mask.}
	\label{fig:rossler_mem}
\end{figure}

We then studied the effect of the memory length and mask type on the photonic NG-RC performance. The R\"ossler observer task was repeated with 5 different initial conditions, formed by randomly choosing $x_1$,  $y_1$, and $z_1$ between 0 to 1  and integrating over 5000 time steps to ensure the system has settled into the chaotic attractor. The average NRMSE is shown in \cref{fig:rossler_mem}(a) as a function of memory length, $K$, for either a unit, decaying, or random mask. For further insight, the top axis displays $t_{\text{R\"ossler},K} = K \Delta t_{\text{R\"ossler}}$ to indicate memory duration in the system time units. The top and bottom ranges of the error bar mark the maximum and minimum NRMSE from the 5 runs, and the data point marks their mean. For each test, we used $N_{\text{train}}=1000$ and $N_{\text{test}}=2000$. For $K<20$ ($t_{\text{R\"ossler},K}<2.5$), the unit memory outperforms both the random and decaying masks. The decaying mask ($\beta=0.8$) experiences an increase in NRMSE for  $K>8$ ($t_{\text{R\"ossler},K}>1$), likely due to a memory-limiting effect as the phase change encoding past history inputs falls below the noise floor. The random mask begins to slightly outperform the unit mask for $K>20$ ($t_{\text{R\"ossler},K}>2.5$), reaching a minimum mean NRMSE of $1.71\times 10^{-2}$ ($\text{NRMSE}_y=1.51 \times 10^{-2}$, $\text{NRMSE}_z=1.94 \times 10^{-2}$) at $K=30$ ($t_{\text{R\"ossler},K}=3.75$). This result improves on the state-of-the-art for R\"ossler observers based on physical RC systems ($\text{NRMSE}_y = 3.6\times 10^{-2},\ \text{NRMSE}_z = 0.1$) \cite{maShortwavelengthReverberantWave2022}. While all-digital RCs have reported lower NRMSE for noise-free data, our result is comparable to the NRMSE obtained in a digital RC with simulated additive noise at an signal-to-noise ratio (SNR) of 40 dB   \cite{luReservoirObserversModelfree2017}. (See SI for more information on photonic NG-RC performance versus SNR). Note that all-digital NG-RCs have not been applied to the R\"ossler task, to the best of our knowledge, so we cannot make a direct performance comparison. We are able to compare more closely in the next section in which the Lorenz system is investigated.

In addition to high prediction accuracy, the photonic NG-RC requires a dramatically shorter warm-up time than traditional physical or 
digital RCs. The photonic NG-RC only requires a warm-up time equal to $K$, the number of memory elements in the input vector. The NG-RC optimized for the R\"ossler system requires only 30 time steps to warm up, compared to physical RCs which often require $10^3$ to $10^5$ steps \cite{luAttractorReconstructionMachine2018,griffithForecastingChaoticSystems2019}. Another major advantage of NG-RCs is that they require fewer training points than conventional RCs. We explore this in depth by studying the NRMSE versus the number of training points $N_{\text{train}}$. The results are shown in \cref{fig:rossler_mem}(b) for a fixed memory of length $K=30$ and a random memory mask. We begin to see satisfactory performance for $N_{\text{train}}$ as low as 200, finding $\text{NRMSE}=5.3\times 10^{-2}$. The errors drop as $N_{\text{train}}$ increases until the improvement tapers off at $N_{\text{train}}=500$ ($\text{NRMSE} = 2.6\times 10^{-2}$). At this point the training duration corresponds to a time interval $t_\text{R\"ossler} = 62.5$, compared to $t_\text{R\"ossler} = 260$ training period for the simulated RC in Ref. \cite{luReservoirObserversModelfree2017}. 
\section{Lorenz System}
\begin{figure*}[!htb]
	\centering
		\includegraphics[scale=\figscale]{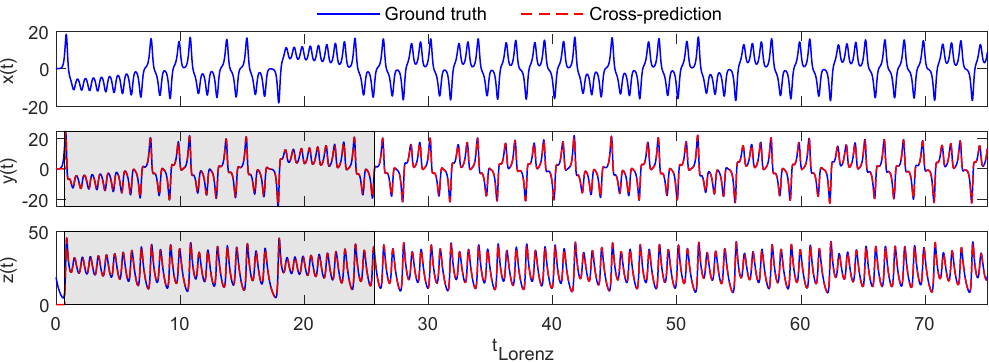}
	\caption{Output of the Photonic NG-RC observer (red dashed) with $K=30$ and a random mask compared to ground truth (blue solid) for $N=3000$ samples of the Lorenz system. The grey shaded region marks the training set of $N_{\text{train}}=1000$ points following a $K-1 = 29$-step warm-up time. The errors for the data shown are $\text{NRMSE}_y= 2.04 \times 10^{-2}$, $\text{NRMSE}_z= 2.77 \times 10^{-2}$.}
	\label{fig:lorenz_single}
\end{figure*}

Having found good results the R\"ossler observer, it is important to verify that the NG-RC applies more generally to a diverse set of problems. In this section, we examine the Lorenz system
\begin{align}
	x_t &= a(y-x) \nonumber \\
	y_t &= x(b - z) -y \nonumber \\
	z_t &= xy - cz,
\end{align}
with the standard parameters $a=10$, $b=28$ and $c=8/3$. This system is also chaotic and characterized by a double-lobe attractor. We again employ the photonic NG-RC as an observer that cross-predicts the $y$ and $z$ coordinates from measurements of $x$. We generate time series by fourth-order Runge-Kutta integration with a step size of $\Delta t_{\text{Lorenz}} = 0.025$. As in the previous section, we begin with a demonstration of results for a well-tuned photonic NG-RC and follow up with the memory hyperparameter investigation that informed the optimization. Finally, we repeat the experiment on time-series data with doubled step size $\Delta t_{\text{Lorenz}}=0.05$ to separate the performance impact of changing $K$ and $t_{\text{Lorenz},K}$. Lorenz cross-prediction results for the parameters  $\Delta t_{\text{Lorenz}}=0.025$, $N_{\text{train}}=1000$, and $N_{\text{test}}=2000$ are shown in \cref{fig:lorenz_single}. The memory is defined to have $K=30$ ($t_{\text{Lorenz},K}=0.75$) and a random mask, and the resulting errors are $\text{NRMSE}_y=1.99\times 10^{-2}$ and $\text{NRMSE}_z=2.47\times 10^{-2}$.

\begin{figure}[!htb]
	\centering
	\includegraphics[scale=\figscale]{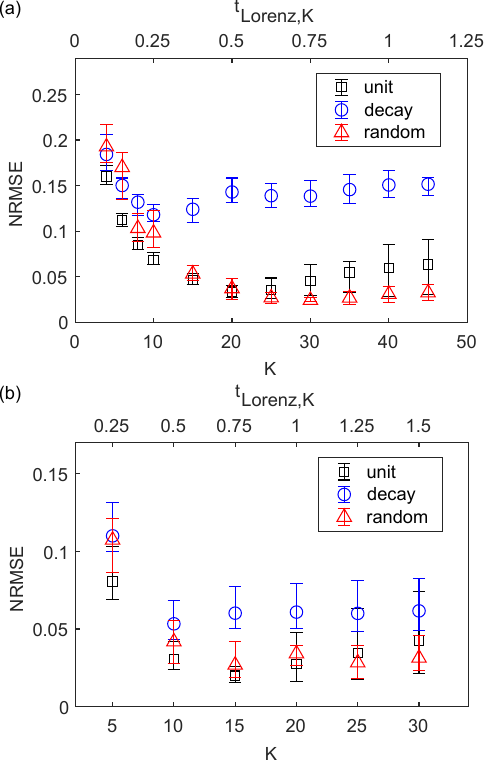}
	\caption{a) Lorenz obsever NRMSE versus memory length $K$/duration $t_{\text{Lorenz},K}$ and mask type, with system step size $\Delta t_{\text{Lorenz}} = 0.025$. b) A repeat of part a with doubled step size $\Delta t_{\text{Lorenz}} = 0.05$}.
	\label{fig:lorenz_mem}
\end{figure}

We studied the prediction accuracy as a function of memory length and mask type, as shown in \cref{fig:lorenz_mem}(a). The error bars are generated the same way as in the previous section. One point of difference with the R\"ossler system is the appearance of a larger separation between the error obtained using unit and random memories. However, the hyperparameter optimization of the Lorenz system shares key features with the R\"ossler equations: the unit mask incurred the smallest error for short memories ($K\leq 20$, $t_{\text{Lorenz},K}<0.5$), the best results occurred for a random mask with $K=30$ ($\text{NRMSE}_y=2.04\times 10^{-2}$, $\text{NRMSE}_z=2.77\times 10^{-2}$), and the worst performance was found for the decaying mask (with $\beta = 0.8$).

Next, we double the time series step size to $\Delta t_{\text{Lorenz}}=0.05$. As seen in \cref{fig:lorenz_mem}(b), the required memory length for peak performance is halved to $K=15$. Since the optimal memory extent $t_{\text{Lorenz},K} =0.75$ is unchanged, this suggests that the memory should be optimized in terms of system time. In contrast to \cref{fig:lorenz_mem}(a), the smallest mean NRMSE of $2.00\times 10^{-2}$  ($\text{NRMSE}_y=1.61\times 10^{-2},\text{NRMSE}_z=2.38\times 10^{-2}$) occurred for a unit memory mask, improving slightly upon the $K=30$ random mask at step size $\Delta t_{\text{Lorenz}}=0.05$. For an overall performance comparison, the digital NG-RC in Ref. \cite{gauthierNextGenerationReservoir2021} achieved $\text{NRMSE}_z=1.75\times 10^{-2}$ for a Lorenz observer using variables $x$ and $y$ to predict $z$. The  results presented here are comparable despite using only $x$ values as input. Finally, we note that the errors for a random memory mask drop below those of the unit mask at $K\geq 25$, indicating that the total number of points in the input feature vector may dictate the optimal choice for the mask function. This suggests that systems with higher dimensionality, even if only short memories are required, may benefit the most from the application of a random weighting of the input vector elements. 

\section{Kuramoto-Sivashinsky system}

In the previous two sections, we showed that the photonic NG-RC  works well for low-dimensional systems. However, one of its largest strengths is the ability to scale to the analysis of high-dimensional systems without a significant increase in power consumption or computational cost. To illustrate this capability, we tested the performance of the fiber NG-RC as an observer for the spatiotemporally chaotic Kuramoto-Sivashinsky (KS) system defined by the partial differential equation
\begin{equation}\label{eq:KS}
	y_t=-yy_x-y_{xx}-y_{xxxx}. 
\end{equation}
The data consist of $P=64$ evenly spaced grid points between spatial coordinates $x=0$ and $x=22$. Choosing $Q$ equidistant spatial samples to be continuously available, the observer interpolates the profile by cross-predicting the other $P-Q$ elements. The ground truth KS data is found by integrating \cref{eq:KS} with the Crank-Nicholson method using periodic boundary conditions and a time step $\Delta t_{\text{KS}}=0.25$. In this case, the NRMSE is defined as \cite{luAttractorReconstructionMachine2018}
\begin{equation}
	\text{NRMSE} = \sqrt{\frac{\sum_{s,n}\left(y_{s,n} - \hat{y}_{s,n}\right)^2}{\sum_{s,n} y_{s,n}^2}},
\end{equation}
computed over all time ($n$) and space ($s$) indices for cross-predicted points during the test phase $n > N_{\text{train}}$. All error bars are computed with the same procedure as in the R\"ossler and Lorenz systems.
\begin{figure*}[!htb]
	\centering
		\includegraphics[scale=\figscale]{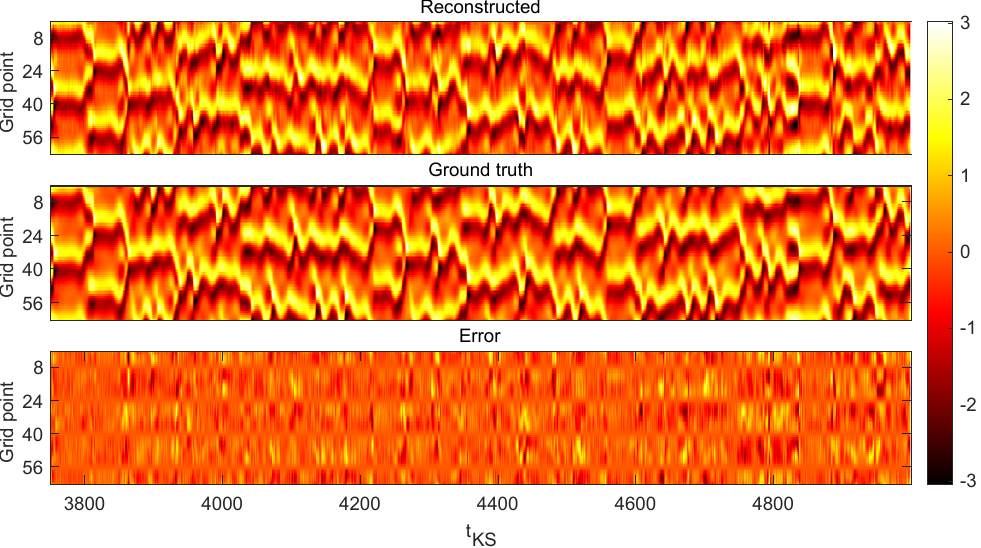}
	\caption{Reconstruction of the spatiotemporal pattern for the KS system using a photonic NG-RC with $K=10$ and a random mask. The input dimension is $Q=4$, with the sample points (8,24,40,56) coinciding with the tick marks on the panel $y$ axes. The top panel shows the reconstruction, the middle contains the ground truth, and the bottom shows the relative error.}
	\label{fig:ks_single}
\end{figure*}
\cref{fig:ks_single} provides an example of an interpolated profile for $Q=4$ observer positions, memory of length $K=10$ ($t_{\text{KS},K}=2.5$), training length $N_{\text{train}}=15000$ ($t_{\text{KS}}=3750$) and testing on the remaining $N_{\text{test}}=5000$ time steps (up to $t_{\text{KS}}=5000$). The top subfigure contains the spatiotemporal pattern recovered by the fiber NG-RC, the middle panel shows the ground truth, and the bottom panel shows the error. We found that the photonic NG-RC successfully reproduced the general spatiotemporal pattern with $\text{NRMSE}=0.30$. 

\begin{figure}[!htb]
	\centering
	\includegraphics[scale=\figscale]{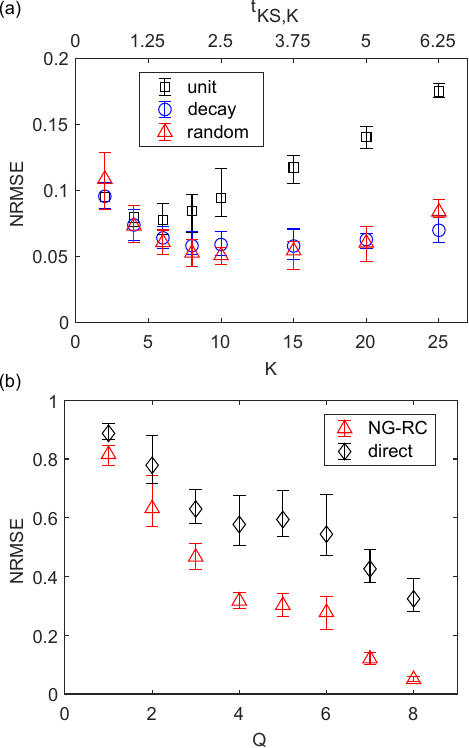}
	\caption{a) NRMSE versus memory length ($K$)/duration ($t_{\text{KS},K}$) and mask function for the KS observer with $Q=8$ observation points. b) NRMSE versus $Q$ for the RC compared to a direct ridge regression on input vectors.}
	\label{fig:ks_mem}
\end{figure}

Next, we increased the input dimensionality to $Q=8$ and repeated the experiment with $N_{\text{train}}=15000$ and $N_{\text{test}}=5000$ while adjusting the memory. As shown in \cref{fig:ks_mem}(a), the error initially decreased with memory length, reaching a minimum $\text{NRMSE}=5.04\times 10^{-2}$ for a random mask with $K=10$ ($t_{\text{KS},K}=2.5$). This value represents the state-of-the-art for RC observers, improving upon the all-digital results in Ref. \cite{luReservoirObserversModelfree2017} while reducing training requirements by a factor of 4. 
In contrast to the R\"ossler and Lorenz systems, we find that both the random and decaying masks lead to lower error than the unit mask for all but the shortest memory ($K=2$, $t_{\text{KS},K}=0.5$). We make sense of this outcome based on the results in figure \cref{fig:lorenz_mem}(b): the most important factor determining the optimal memory mask is the total number of points in the input feature vector. For example, at $Q=8$ and $K=4$, the input feature vector length is 32. For the R\"ossler and Lorenz system, the random mask outperformed the unit mask for feature vectors with $K\geq 25$ ($Q=1$), or a total input feature vector size of 25. Thus, for high-dimensional systems such as the KS system, the random mask provides superior performance. We also note that the random mask was capable of optimal or near-optimal performance for each of the systems studied here, making it a good initial option when applying the photonic NG-RC to new tasks. 

Finally, we measured the NRMSE versus input dimension $Q$ using the optimized memory of length $K=10$ and a random mask. To evaluate the importance of generating output feature vectors by a nonlinear transformation, we also attempted to predict the KS pattern by applying a ridge regression directly to the input feature vectors, $\mathbf{v}_{\text{in},n}$. As shown in \cref{fig:ks_mem}(b), the NG-RC outperformed this linear optimization for each value of $Q$, confirming the importance of applying a nonlinear transform to the input data. In comparison to results reported for digital RCs on the KS task \cite{luReservoirObserversModelfree2017}, the photonic NG-RC achieved similar NRMSE for $Q\leq6$ and improved on the state-of-the-art for $Q\geq7$. Again, these results were obtained using significantly less training data and without the long warm-up time required in traditional RCs.

\section{Energy consumption and latency}

In addition to providing state-of-the-art predictive performance, the photonic NG-RC has the potential for significantly lower latency than digital electronic (software-based) NG-RCs. Accounting for the averaging required to increase SNR, the photonic NG-RC presented in this work requires $\sim$125 $\mu$s (R\"ossler, Lorenz) or $\sim$25 $\mu$s (KS) to map an input feature vector to an output feature vector. We may improve this transformation time by implementing two changes. The first is to increase the signal power to eliminate the need for averaging. The second is to increase the encoding rate from the 100 MHz used in this work to 10 GHz, a typical modulation rate in optical fiber. As a result, we reduce the time to generate output feature vectors to 5 ns using a 5 m fiber. To eliminate the latency due to output vector digitization and software-based weighting, we may apply output weights instead by the analog scheme from Refs. \cite{duportFullyAnaloguePhotonic2016,smerieriAnalogReadoutOptical2012} to complete inference with a total latency of 5 ns. The photonic NG-RC could also support multiplicative increases in speed through spatial or wavelength multiplexing techniques, providing the potential for exceptionally low latency.

We also analyzed the energy consumption required by the photonic NG-RC to map an input feature vector to an output feature vector. In particular, we considered the power consumption required by the laser, the modulators, the photodetector, the digital to analog converter (DAC) used to drive the modulator and the analog to digital converter (ADC) used to record the RBS pattern. Details of the energy consumption calculation are provided in the SI. We then compared the photonic NG-RC to the energy consumption required by a digital electronic NG-RC \cite{gauthierNextGenerationReservoir2021}, which constructs output feature vectors by applying nonlinear transformations in software. For this comparison, we first calculated the number of multiplication operations required to generate a single output feature vector. We compute this number as a function of the dimensionality of the observed data, $Q$, the memory length, $K$, and the highest monomial order of the output feature vector, $d$. The number of multiplication operations required to generate feature vectors of the form described in Ref. \cite{gauthierNextGenerationReservoir2021} can be expressed as: 
\begin{equation}
	M_{\text{\text{NG-RC}}} = \sum_{p=2}^d \binom{KQ + p - 1}{p} - \binom{(K-1)Q + p - 1}{p},
\end{equation}
where the parentheses denote the binomial coefficient operator. The detailed derivation of $M_{\text{NG-RC}}$ is provided in the SI. Assuming that each multiplication required $\approx$ 0.5 pJ (which is typical of state-of-the-art GPUs, as well as other digital electronic computing architectures \cite{wangIntegratedPhotonicEncoder2023}), the energy required to generate the output feature vector at each time step can be expressed as $E_{\text{NG-RC}}=M_{\text{NG-RC}}\times0.5$ pJ. Note that this analysis can be considered a conservative estimate because it did not account for the power required to store and access memory. In contrast, the photonic NG-RC effectively performs “in-memory” computing and does not need to store intermediate values to compute the output from a given input vector. 

\begin{figure}[!htb]
	\centering
	\includegraphics[scale=\figscale]{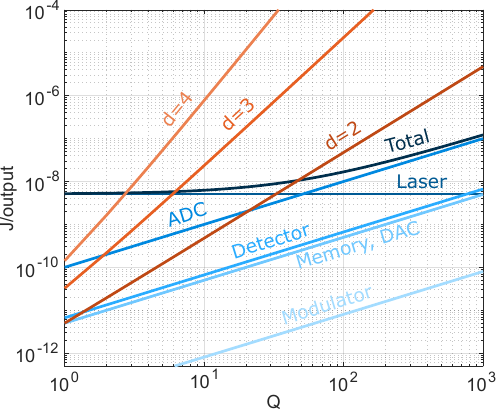}
	\caption{Energy required to generate an output feature vector, $\mathbf{v}_{\text{out},n}$, for the photonic NG-RC (blue lines) and the digital electronic NG-RC described in Ref. \cite{gauthierNextGenerationReservoir2021} (orange lines). The individual blue curves indicate the energy required by varying components and the dark blue line labelled “Photonic NG-RC” represents the total energy consumption. The memory and DAC operations have identical energy contribution curves. The photonic scheme requires lower energy for systems with large $Q$ or high non-linearity (requiring large $d$). }
	\label{fig:energy}
\end{figure}

To compare the energy consumption of the photonic NG-RC with that of a monomial-based digital-electronic NG-RC, we first note that the nonlinear transformation imposed by the photonic system is different from the explicit construction of monomials used to form output feature vectors in the digital approach. However, using the results of this work, we assume that the performance of a monomial-based NG-RC is equivalent to a photonic NG-RC with output feature vector length equal to 10 times the input vector length. \Cref{fig:energy} compares the energy consumed by equivalent-performance photonic and software-based NG-RCs as a function of the input feature vector length $(KQ)$ for a fixed memory of length $K=10$. In the photonic approach, we found that the laser is the primary source of energy consumption for $Q<100$. The ADC dominates at larger $Q$ as the number of required operations increases with the size of the output feature vectors. For the digital-electronic NG-RC, the energy requirements increase with $Q$ in all cases. Repeating for digital NG-RCs with highest order monomial $d = 2$, 3, or 4, we find that increasing $d$ increases the slope in energy consumption as a function of $Q$. We find that the monomial-based NG-RC is more energy efficient than the photonic NG-RC For low-dimensional systems with modest nonlinearity (which can be captured using $d=2$ or 3). However, as the input dimensionality increases, particularly in systems requiring higher-order nonlinearities, the photonic NG-RC provides the potential for dramatically lower energy consumption. For example, with input dimension $Q=100$ and $d=3$, the photonic NG-RC could generate an output feature vector using 1000$\times$ less energy than the software-based implementation. It is possible to further reduce the energy requirements by applying analog weights at the prediction phase \cite{duportFullyAnaloguePhotonic2016,smerieriAnalogReadoutOptical2012}, so that only a single ADC operation would be required to obtain a prediction.
\section{Discussion}
In this work, we introduced a photonic NG-RC that uses Rayleigh backscattering to implement a nonlinear map between input and output feature vectors using a combination of coherent, interferometric mixing and a quadratic readout. We showed that this type of nonlinear projection enables excellent performance on a variety of benchmark tasks while maintaining the advantages of digital electronic NG-RCs: short warm-up time, reduced training time, and simple hyper-parameter optimization. The photonic NG-RC also addresses the scalability limitations of digital electronic NG-RCs. Since the RBS process is entirely passive, the photonic NG-RC is capable of performing the equivalent of $KQD$ multiply and accumulate (MAC) operations while requiring only $KQ$ modulations (to encode the input feature vectors). As a result, the photonic NG-RC has the potential for dramatically lower latency and power consumption than a digital-electronic NG-RC when analyzing high-dimensional and/or highly nonlinear systems.

There are also several areas for improvement. First, in this demonstration, we used a data encoding rate of 100 MHz, leading to latency of 5 $\mu$s in the generation of output feature vectors. The effective latency was further increased by a factor of 5 (KS) or 25 (R\"ossler, Lorenz) due to the averaging employed to increase SNR. In the future, higher-speed operation could enable similar computations in as little as 5 ns. We may also use spatial or wavelength multiplexing to further accelerate the photonic NG-RC. Second, NG-RC performance depends strongly on the ability of the nonlinear transformation to capture the physics of the dynamical system being studied. Fortunately, additional nonlinearities are easily accessible in this platform with minimal impact on power consumption or latency. Highly nonlinear fiber could be used to access all-optical nonlinearities in the RBS fiber while maintaining modest power levels. It is also possible to introduce additional nonlinearity in the readout stage by choosing a photo-detector with lower saturation power to clamp the measured RBS signal. We expect that the optimal transformation will depend strongly on the target task. Finally, in this demonstration, the entire input feature vector was encoded on an optical carrier at each time step. In the future, delayed copies of the observed data could be created using a series of delay lines, requiring the observed data at a given time step to be encoded only once. Although \cref{fig:energy} showed that the re-encoding cost is relatively modest (see the energy consumed by the DAC and modulator), such an approach could enable the photonic NG-RC to operate directly on data transmitted over fiber in the analog domain without requiring intermediate digitization and storage.  

In summary, this work showed that a photonic platform can efficiently apply the nonlinear transformation required for NG-RC simply by performing random nonlinear projections. While the fiber backscattering architecture presented here has many attractive features, we expect that this work may spur the development of alternative photonic NG-RC platforms tailored to different applications that take advantage of the design flexibility afforded by foregoing the need to construct a physical cavity. 

\section*{Supplementary Material}
See the supplementary material for a mathematical model of the photonic NG-RC system, an analysis on the effect of averaging on performance, and a derivation of the power consumption for photonic and digital electronic NG-RCs.

\begin{acknowledgments}
The authors acknowledge support from the U.S. Naval Research Laboratory (6.1 Base Funding).
\end{acknowledgments}
\section*{Conflicts of Interest}
The authors have no conflicts to disclose.

\section*{Data Availability Statement}
The data that support the findings of this study are available from the corresponding author upon reasonable request.

\bibliography{RC_bib}

\end{document}


	
	\title{Photonic next-generation reservoir computer based on distributed feedback in optical fiber: Supplementary Information}
	
	\author{Nicholas Cox}
	\email{nicholas.a.cox34.ctr@us.navy.mil} 
	\author{Joseph Murray}
	\author{Joseph Hart}
	\author{Brandon Redding}
	
	\affiliation{U.S. Naval Research Laboratory, 4555 Overlook Ave, SW, Washington, DC 20375, USA}
	
	\date{\today}
	
	\maketitle
	\section{Model}
	Here, we present a physical model that describes the action of the photonic NG-RC. We encode the data set into the phase of an optical pulse by first generating the $\phi$-EOM drive voltage
	\begin{equation}\label{eq:V}
		V_{\text{in}}(t) = \sum_{n = 1}^N V_{\text{in},n}(t - nT),
	\end{equation}
	in which input vectors at each time step are encoded into $V_{\text{in},n}(t)$ separated by period $T$. Each $V_{\text{in},n}(t)$ takes the form
	\begin{equation}\label{eq:Vn}
		V_{\text{in},n}(t)= V_0 + \frac{1}{V_s}\sum_{\mu=1}^{KQ} v_{\text{in},n}^\mu \text{rect} \left(\frac{t-\mu\tau}{\tau}\right),
	\end{equation}
	where $\tau$ is the temporal width of a single data point. $V_0$ and $V_s$ offset and scale the signal so that the maximum and minimum voltages of $V_{\text{in}}(t)$ (\cref{eq:V}) are $\pm V_{\pi}$, respectively. The value $v_{\text{in},n}^\mu$ is the $\mu$ component of the length-$KQ$ input vector with a memory mask applied. Driving the $\phi$-EOM with the voltage from \cref{eq:Vn} generates the phase envelope 
	\begin{equation}\label{eq:phip}
		\phi_n(t)=\pi V_{\text{in},n}(t)/V_\pi
	\end{equation}
	at time step $n$. After applying the phase shift $\phi_n(t)$ to the laser source, we use an acousto-optic modulator to carve a pulse of width $W = (KQ + 6)\tau$ to yield the input electric field
	\begin{equation}\label{eq:Ein}
		E_{\text{in},n}(t) = |E_0|\text{rect}\left(\frac{t - W_0}{W}\right)e^{-i(\omega t - \phi_{n}(t))}
	\end{equation}
	at time $n$. The time offset argument $W_0 = (KQ + 5)\tau/2$ simply ensures that the amplitude modulation is centered with the extent of the phase-modulated signal. The optical pulse in \cref{eq:Ein} is then sent through a single-mode fiber of length $L$ which creates a backward-moving output field via Rayleigh backscattering (RBS). 
	
	We compute the RBS response in the frequency domain by first decomposing the electric field into plane wave components $E(\omega) = \mathscr{F}\{E_{\text{in},n}(t)\}$ (where $\mathscr{F}$ is the Fourier transform operator). We then continue to define the field reflection coefficient for each frequency component. Fix $z = 0$ to be the beginning of the scattering fiber. A scattering site at position $0 < z_j \leq L$ reflects a plane wave $e^{i(k(\omega)z - \omega t)}$ at angular frequency $\omega$ with complex coefficient $r_j e^{2 i k(\omega) z_j}$, where $k(\omega)$ is the optical dispersion relation. The exponent represents the total phase accumulated as the plane wave travels to the scattering site and back to the origin. Combining a total of $N_r$ reflectors, we find the total reflection coefficient in the frequency domain to be
	\begin{equation}\label{eq:rtot}
		r_{\text{RBS}}(\omega) = \sum_j^{N_r} r_j \exp\left({- 2 i \frac{n(\omega)\omega}{c} z_j}\right).
	\end{equation}
	In \cref{eq:rtot}, we used the fact that $k(\omega) = n(\omega)\omega/c$, and calculate the value of $n(\omega)$ from the appropriate Sellmeier coefficients neglecting waveguide dispersion. The RBS-scattered field is then calculated by applying the reflection coefficient to each input frequency component and taking the inverse Fourier transform to convert back to the time domain:
	\begin{equation}\label{eq:ERBS}
		E_{\text{RBS},n}(t) = \mathscr{F}^{-1}\{E_{\text{in},n}(\omega)r_{\text{RBS}(\omega)}\}.
	\end{equation}
	By the convolution theorem, \cref{eq:ERBS} is identical to the expression  
	\begin{equation}\label{eq:conv}
		E_{\text{RBS},n}(t) = E_{\text{in},n}(t)*r_{\text{RBS}}(t),
	\end{equation}
	for the time-domain filter $r_{\text{RBS}}(t) = \mathscr{F}^{-1}\{r_{\text{RBS}}(\omega)\}$. Finally, the detection process generates a voltage proportional to the RBS irradiance
	\begin{equation}\label{eq:Vout}
		V_{\text{out},n} \propto |E_{\text{RBS},n}(t)|^2.
	\end{equation}
	Once digitized, the evenly-sampled $V_{\text{out},n}$ serves as the output feature vector $V_{\text{out},n}$ to which we apply the final output weights to make predictions. As seen in Eq. (3) of the main paper, we can determine the net nonlinear transformation from input to output vectors by observing the voltage relationship formed by \cref{eq:Vn,eq:phip,eq:Ein,eq:conv,eq:Vout}.
	\section{Effect of averaging on NRMSE}
\begin{figure}[h!]
		\centering
		\includegraphics[scale=1.0]{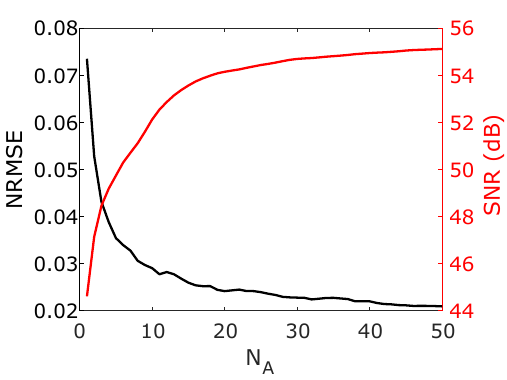}
		\caption{NRMSE and SNR versus number of averaged repetitions}
		\label{fig:rossler_avg}
	\end{figure}

	To understand the relationship between noise and NRMSE, we passed the same R\"ossler input data with $K=30$ through the photonic NG-RC 50 times. Averaging $N_A$ RBS patterns and computing the mean NRMSE for each value of $N_A$, we generated the data shown in the left axis of \cref{fig:rossler_avg}. For each $N_A$, we also calculated the SNR by
	\begin{equation}\label{eq:snr}
		\text{SNR (dB)} = 20 \log\left(\frac{\overline{V}_{\text{rms}}}{\sigma_b}\right),
	\end{equation}
	where $\overline{V}_{\text{rms}}$ is the measured root-mean-square voltage with DC offset removed and $\sigma_b$ is the standard deviation of the measured background signal. 
	We find rapid improvement in the NRMSE as we increase from a single shot to a 10-run average, corresponding to an increase in SNR from 45 dB to 52 dB. In this work, the R\"ossler and Lorenz systems are studied by averaging 25 repetitions for an SNR of 55 dB. Because the KS system requires longer training sets, we are limited by data acquisition hardware to a 5 repetition average (SNR = 48 dB) for this task.
	We find context for the noise behavior by comparing to the simulations  in Ref. \cite{luReservoirObserversModelfree2017}. The authors study the effect of observational noise by adding a random number to each input data point. These random numbers are chosen from a uniform distribution between -$\eta$ and $\eta$ for some magnitude $\eta$. Because the input signals are normalized to have zero mean and unit variance, the input SNR is found from \cref{eq:snr} to be
	\begin{equation}\label{eq:digital_noise}
		\text{SNR (dB)} = -20 \log \left(\sqrt{3} \eta \right).
	\end{equation}
	Because the reservoir in Ref. \cite{luReservoirObserversModelfree2017} is entirely digital, the total noise is assumed to be represented by \cref{eq:digital_noise}. For observation of $y$ variable, they found $\text{NRMSE}_y = 5.2\times 10^{-3}$ at SNR = 55 dB ($\eta=10^{-3}$) and $\text{NRMSE}_y=1.9\times 10^{-2}$ at SNR = 35 dB ($\eta=10^{-2}$). In the main paper, we find the best performance $\text{NRMSE}_y = 1.2 \times 10^{-2}$ for $N_A = 25$ (SNR = 54 dB). Accordingly, we find that our experimental system performs slightly worse at the same SNR. This difference may be partially explaiend by environmental drift in the photonic NG-RC.
	
	\section{NG-RC Energy Consumption}
	This section compares the energy consumption of the photonic NG-RC to that of a digital-electronic NG-RC. We define the energy required to be the amount needed to generate a single output feature vector $\mathbf{v}_{\text{out},n}$ from an input vector $\mathbf{v}_{\text{in},n}$.
	
	\subsection{Energy consumption in the photonic NG-RC}
	In  this section, we estimate the energy required by the photonic NG-RC to map a linear input feature vector $\mathbf{v}_{\text{in}}$ of length $KQ$ to an output feature vector $\mathbf{v}_{\text{out}}$ of length $D$. Our analysis includes the energy required to operate the laser, modulators, photodetector, digital-to-analog converter (DAC), analog-to-digital converter (ADC), and the energy required to digitally generate the input feature vector (i.e. the energy required to apply the memory mask). We adapt this analysis from the one detailed in Ref. \cite{reddingFiberOpticComputing2024}.
	We first estimate the optical power required at the detector to achieve a desired signal-to-noise ratio (SNR), assuming shot-noise limited detection as:
	\begin{equation}
		P_{\text{Rx}}= \frac{2 q f_0}{R}\cdot \text{SNR} ,
	\end{equation}
	where $\text{SNR} = 10^{\text{SNR (dB)}/10}$, $q$ is the charge of an electron, $f_0$ is the measurement bandwidth, and $R$ is the responsivity of the detector. We then estimate the required laser power as
	\begin{equation}
		P_\text{laser} = P_{\text{Rx}}/\left(T_{\text{mod}} r_{\text{RBS}}\right),
	\end{equation}
	where $T_\text{mod}$ is the transmission through the modulators and $r_\text{RBS}$ is the average reflectance due to Rayleigh backscattering in the fiber. In this analysis, we assumed the standard single-mode fiber is replaced with enhanced Rayleigh backscattering fiber with reflectance \cite{handerekImprovedOpticalPower2018}
	\begin{equation}
		r_{\text{RBS}}= (-67 \text{ dB/ns})\cdot (KQ + 6)\tau,
	\end{equation}
	where $(KQ + 6)\tau$ is the total input optical pulse duration including the buffer $6 \tau$.  The total electrical power required to operate the laser is then $P_{\text{laser}}/\eta$, where $\eta$ is the wall-plug efficiency. Assuming the laser is active only during the input encoding process, we can express the laser energy required to generate a single $\mathbf{v}_{\text{out},n}$ as
	\begin{equation}
		E_{\text{laser}}= \frac{P_{\text{laser}}}{\eta} (KQ+6)\tau.
	\end{equation}
	The energy consumed by the modulator to encode $\mathbf{v}_{\text{in}}$ is calculated by \cite{millerEnergyConsumptionOptical2012}
	\begin{equation}
		E_{\text{mod}}= \frac{1}{2} KQ C_{\text{mod}} V_\pi^2,
	\end{equation}
	where $C_{\text{mod}}$ is the capacitance of the modulator and $V_\pi$ is the modulator voltage required to impart a $\pi$ phase shift. The energy consumed by the photodetector during a measurement of $\mathbf{v}_{\text{out}}$ can be expressed as 
	\begin{equation}
		E_{\text{det}} \approx V_{\text{bias}} R P_{\text{Rx}} (D/f_0 ),			
	\end{equation}
	where $V_{\text{bias}}$ is the bias voltage applied to the detector.
	State-of-the-art DACs and ADCs require $\sim$ 0.5 pJ/use and $\sim$ 1 pJ/use, respectively \cite{wadaGHzbandCMOSDirect2012,caragiuloDACPerformSurv,murmannADCPerformSurv}. The DAC is used each time we switch the modulator, which occurs $KQ$ times to encode $\mathbf{v}_{\text{in}}$, whereas the ADC is used to record all $D$ samples contained in $\mathbf{v}_{\text{out}}$, resulting in energy consumption of
	\begin{align}
		E_{\text{DAC}}&=KQ \times 0.5 \text{ pJ} \\					
		E_{\text{ADC}}&=D \times 1 \text{ pJ}.					
	\end{align}
	Finally, the application of a non-unity memory mask requires $KQ$ multiplications to be performed digitally. This mask may be implemented more efficiently in hardware, but as we will see, the energy consumed to apply the mask is negligible. Assuming 0.5 pJ per multiplication (typical for a GPU \cite{wangIntegratedPhotonicEncoder2023}), there is additional energy consumed to apply the mask of
	\begin{equation}
		E_{\text{Memory}}= KQ \times 0.5 \text{ pJ}.
	\end{equation}
	We then find the total energy consumption by summing that of the laser, modulator, detector, DAC, ADC, and mask function.
	
	To estimate the energy required by the photonic NG-RC, we first assumed $D=10KQ$, corresponding to expanding the dimensionality of the input vector by a factor of 10. This is comparable to the degree of dimensionality expansion used in this work (e.g. with $D \approx 500$, $K=30$, and $Q=1$ for the R\"ossler and Lorenz systems). We then set the required measurement SNR to 48 dB (per \cref{fig:rossler_avg}, similar to the experimental SNR used in the measurements reported in this work after averaging 5 sequential measurements). The remaining parameters were set to typical values: a laser with wall-plug efficiency of $\eta=0.2$, modulators with $C_\text{mod} =1$  fF, $V_\pi= 4$ V, and 2 dB insertion loss ($T_\text{mod}=0.40$ for both modulators), and a detector with $V_\text{bias} = 3$ V and $R=1$ A/W \cite{nozakiFemtofaradOptoelectronicIntegration2019,li25Gb1VdrivingCMOS2011}. Using these expressions, we were able to estimate the energy that an optimized photonic NG-RC would consume as a function of the size of input dimensionality $Q$. As shown in Fig. 9 in the main manuscript, the photonic NG-RC has the potential for substantially lower energy consumption than a software-based NG-RC for large $Q$ and/or systems requiring complex nonlinearities. 
	\subsection{Energy consumption in a digital electronic (software-based) NG-RC}
	Digital electronic (software-based) NG-RCs create an output feature vector from $K$ time-delayed copies of the $Q$ observed variables. Here, we assume the standard form \cite{gauthierNextGenerationReservoir2021} of the output feature vector containing monomials up to order $d$ as the nonlinear functions. Including the constant term 1, all output feature vector elements are formed by multiplying $d$ elements of the set of lowest-order terms and retaining the unique values. Beginning with $KQ+1$ zero- and first-order terms, a feature vector with order $d$ monomials has length $_{KQ+1}C^{(\text{R})}_{d}$, where $_n C^{(\text{R})}_k $ denotes the number of ways to select $k$ items from a set of size $n$ with replacement. Mathematically, this quantity is given by
	\begin{equation}
		_n C_k^{(\text{R})} = \binom{n+k-1}{k},
	\end{equation}
	where
	\begin{equation}
		\binom{n}{k} = \frac{n!}{k!(n-k)!}.
	\end{equation}
	An example feature vector, with parameters $K=3$, $Q=1$, and $d=2$, would be:                      
	\begin{equation}
		\mathbf{v}_{\text{out},n} = \begin{bmatrix}
			1 \\ x_n \\ x_{n - 1} \\ x_{n-2} \\ x_n^2 \\ x_{n-1}^2 \\ x_{n-2}^2 \\ x_n x_{n-1} \\ x_{n} x_{n-2} \\ x_{n-1}x_{n-2}
		\end{bmatrix},
	\end{equation}
	where $x$ is the observed variable provided as input to the NG-RC. 
	To estimate the energy consumption required to generate $\mathbf{v}_{\text{out},n}$, we determine the number of unique multiplication operations required at the $n^{\text{th}}$ time-step. We will proceed order-by-order. Both the constant and first-degree monomials require no multiplication, so they do not increase the energy cost of feature vector generation. Begin then with order $p = 2$: the total number of second-order terms is equal to the number of unique combinations of two first-order terms, $_{KQ}C_2^{(\text{R})}$. Because each of these elements is formed by a single multiplication operation, we may expect $_{KQ}C_2^{(\text{R})}$ operations are needed.  However, we note that elements composed of two previous samples (for example $x_{n-1}x_{n-2}$) will have already been computed at the previous time step $n - 1$. The total number of second-order monomial terms that do not contain information from the current time step is $_{(K-1)Q}C_2^{(\text{R})}$, reflecting the removal of the $Q$ elements at time step $n$ from the combinations. Since this value represents the number of terms carried over from a previous $\mathbf{v}_{\text{out}}$, the number of multiplications needed is equal to the difference
	\begin{equation}\label{eq:M2}
		M_{\text{NG-RC}}^{(2)} = \binom{KQ + 1}{2} - \binom{K(Q-1) + 1}{2}.
	\end{equation} 
	Continuing to the next order, we can generate each third-degree monomial by a single multiplication with a second-degree monomial. Following the same logic as before, we now find a total of $_{KQ}C_3^{(\text{R})}$ terms with $_{K(Q-1)}C_3^{(\text{R})}$ in memory from a previous step. Comparing to \cref{eq:M2}, we extend this pattern to find the total number of multiplications required to form a feature vector with highest monomial order $d$:
	\begin{equation}
		M_{\text{NG-RC}} = \sum_{p=2}^d \binom{KQ + p - 1}{n} - \binom{(K-1)Q + p - 1}{p}.
	\end{equation}
	In order to estimate the energy required to generate the output feature vector, we then assume each multiplication requires $\sim$ 0.5 pJ (typical for GPUs), and thus $E_{\text{NG-RC}} = M_\text{NG-RC}\times 0.5$ pJ.
	\bibliography{RC_bib}